\documentclass[%
 aip,
 amsmath,amssymb,
preprint,%
]{revtex4-1}

\usepackage{graphicx}
\usepackage{dcolumn}
\usepackage{bm}

\usepackage[utf8]{inputenc}
\usepackage[T1]{fontenc}
\usepackage{mathptmx}
\usepackage{xcolor}
\usepackage{color}

\usepackage{subfig}
\usepackage{caption}
\usepackage{tabularx}
\usepackage{soul}

\begin{document}

\preprint{}


	\title[Fibre-based single photon source using an InAsP quantum dot and GRIN lenses]{Optical fibre-based (plug-and-play) single photon source using InAsP quantum dot nanowires and gradient-index lens collection}
	
\author{David B. Northeast}
\email{david.northeast@nrc-cnrc.gc.ca}
\affiliation{National Research Council Canada, Ottawa, Ontario, Canada, K1A 0R6.}

\author{Dan Dalacu}
\affiliation{National Research Council Canada, Ottawa, Ontario, Canada, K1A 0R6.}

\author{John F. Weber}
\affiliation{National Research Council Canada, Ottawa, Ontario, Canada, K1A 0R6.}
\author{Jason Phoenix}
\affiliation{National Research Council Canada, Ottawa, Ontario, Canada, K1A 0R6.}
\affiliation{University of Waterloo, Waterloo, N2L 3G1, Canada}

\author{Jean Lapointe}
\affiliation{National Research Council Canada, Ottawa, Ontario, Canada, K1A 0R6.}

\author{Geof C. Aers}
\affiliation{National Research Council Canada, Ottawa, Ontario, Canada, K1A 0R6.}

\author{Philip J. Poole}
\affiliation{National Research Council Canada, Ottawa, Ontario, Canada, K1A 0R6.}

\author{Robin L. Williams}
\affiliation{National Research Council Canada, Ottawa, Ontario, Canada, K1A 0R6.}

	\date{\today}
	
	\begin{abstract}
		
		We present a compact, fibre-coupled single photon source using gradient-index (GRIN) lenses and an InAsP semiconductor quantum dot embedded within an InP photonic nanowire waveguide. A GRIN lens assembly is used to collect photons close to the tip of the nanowire, coupling the light immediately into a single mode optical fibre. The system provides a stable, high brightness source of fibre-coupled single photons. Using pulsed excitation, we demonstrate on-demand operation with a single photon purity of 98.5\% when exciting at saturation in a device with a source-fibre collection efficiency of 35\% and an overall single photon collection efficiency of 10\%. We also demonstrate ``plug and play" operation using room temperature photoluminescence from the InP nanowire for room temperature alignment. 
		
	\end{abstract}

\maketitle
	
Non-classical photon states (e.g. single, indistinguishable, or entangled) form the basis of photonic quantum technologies\cite{Obrien_NP2009}. For implementation of these technologies in a future quantum network, the efficient transmission of these states between nodes in the network is critical. The most convenient transmission channel is an optical fibre, and in particular, a single-mode fibre (SMF) where photons exist in a single spatial mode. Solid-state two-level systems (e.g. quantum dots, or defects) can produce the required states efficiently and deterministically, but the devices are typically designed to emit into free-space. High efficiency coupling of the free-space mode to the gaussian mode of a SMF can be achieved with external optics\cite{Gazzano_PRL2013,Bulgarini_NL2014,Lee_NL2021} and, using nearly perfectly matched modes, coupling efficiencies of 93\% have been demonstrated\cite{Bulgarini_NL2014}. To eliminate stability issues related to the alignment of free-space optics, new techniques for coupling to fibres are required. One can, for example, use the cleaved facet of a fibre to form one mirror of a Bragg reflector-based open microcavity allowing for direct collection of the cavity photons into the fibre\cite{Muller_APL2009,Greuter_PRB2015}. Alternatively, evanescent coupling via tapered fibres can be applied to collect light from emitters\cite{Fujiwara_NL2011,Liebermeister_APL2014}, cavities\cite{Ates_SR2013,Lee_SR2015} or tapered nanobeam waveguides\cite{Patel_LIGHT2016,Daveau_OPT2017,Burek_PRA2017}.  Although near-unity coupling efficiencies are predicted, measured efficiencies remain substantially below those obtained using free-space optics. 
	
	




The long term goal of fibre-based quantum light source development is a fixed-alignment device in a compact and robust package that can be operated in a turn-key manner. Many techniques have been proposed to eliminate user alignment by permanently attaching a fibre to the source; a so-called fibre-based plug and play (PnP) source. One PnP approach involves positioning a fibre relative to an emitter using various alignment techniques, followed by gluing the fibre in place with epoxy. Initial experiments did not implement deterministic alignment, relying instead on a fibre bundle or array simply pressed to a sample containing emitters, e.g. planar dots\cite{Xu_APL2007} or dots in weak cavity micropillars\cite{Xu_APL2008,Ma_APL2017} and measuring each fibre until one that is aligned to an emitter is found. An interferometric technique for aligning to the micropillar can be empoyed for deterministic coupling \cite{Zolnacz_OE2019} but ultimate efficiencies are only 18.6\% due to the mode mismatch between the pillar and low numerical aperture (NA) SMF fibres. Slightly higher ultimate efficiencies are predicted using microlenses etched around pre-selected quantum dots\cite{Schlehahn_SR2018}. In this case, the fibre is aligned using the host semiconductor photoluminescence from a gold aperture lithographically patterned around the microlens using alignment marks. To further increase the spatial overlap with the fibre mode, a miniaturized double-lens system is patterned around the microlens using a 3D, two-photon direct laser system, complete with a chuck for fibre placement\cite{Bremer_APLP2020}. Although relatively high coupling was achieved between the light emitted from the microlens and the fibre, the overall single photon collection efficiency was low (0.56\%) and attributed to a misalignment of the dot to the microlens. Finally, for devices employing high Q cavities\cite{Haupt_APL2010,Snijders_PRA2018}, active alignment between the fibre and cavity is possible at room temperature by measuring the transmission through or reflection from the cavity using two fibres, one on each side of the chip, which are subsequently glued in place. Measured cavity-fibre coupling efficiencies are high at 85\% and close to the predicted values of 90\%. Measured single-photon collection efficiencies, however, were lower by a factor of 17 due to misalignment of the quantum dot within the cavity\cite{Snijders_PRA2018}.


A second approach relies on positioning an emitter on the facet of a cleaved fiber, where various techniques can be used to align to the fibre core\cite{Schroder_NL2011,Vogl_JPD2017,Sasakura_APE2013,Zha_NT2015}. This near-field coupling approach may have potential with defect-type sources e.g. nitrogen-vacancies in diamond nanocrystals\cite{Schroder_NL2011} or defects in boron nitride flakes\cite{Vogl_JPD2017}, but is not well suited to epitaxial quantum dots\cite{Sasakura_APE2013,Zha_NT2015} which are embedded in a host semiconductor. A more sophisticated approach to attaching a quantum dot-based source to a fibre facet relies on nanowire structures\cite{Cadeddu_APL2016} where the nanowire geometry is used to improve mode-matching. Alternatively, instead of a cleaved facet, tapered fibre tips can used\cite{Lee_APL2019}. In this case, the emitter is located within a tapered nanobeam waveguide which is removed from the growth substrate and placed on the fibre tip, where it couples to the fibre evanescently, in a similar manner to the tapered fibre approaches described above. 


In Table~\ref{PnP_sources} we summarize the performance of sources that have been operated in a PnP mode. From the table, it is clear that several approaches can provide good overlap between the optical modes of the source and the fibre, with predicted coupling efficiencies approaching 100\%. In practice, however, devices dramatically under-perform. Calculated efficiencies typically assume perfect alignment of the fibre and source and ideally this would be done actively using emission from the emitter. Many emitters, however, are insufficiently bright at room temperature (where epoxies are applied) to allow for active alignment to the emitter emission. Instead, alignment is performed either using apertures containing the emitter\cite{Schlehahn_SR2018} or using photonic structures within which the ideal positioning of the emitter is not guaranteed\cite{Snijders_PRA2018,Zolnacz_OE2019,Bremer_APLP2020}.

\begin{table}
\caption{Quantum light sources operated in PnP mode.\label{PnP_sources}}
\begin{tabular}{lcccc}
\hline
\hline
Material system    & Alignment       & $\eta_{\mathrm{calculated}}$      & $\eta_{\mathrm{measured}} $   & Ref.\\
\hline

  QD-$\mu$pillar  &     fibre bundle        &       0.69\%       &   0.095\%    &   \citenum{Xu_APL2007}   \\
  QD-$\mu$pillar  &     fibre array        &       26\%       &   ----    &   \citenum{Ma_APL2017}   \\
   QD-$\mu$pillar  &     reflection  &      18.6\%       &   ----    &   \citenum{Zolnacz_OE2019}   \\
   QD-nanowire     	&    visual       &      9.2\%        &   5.8\%     &   \citenum{Cadeddu_APL2016}   \\
  QD-nanowire     	&    QD PL       &      ----        &   ----     &   \citenum{Zha_NT2015}   \\        	
  QD-flake     &    random       &      ----        &   ----     &   \citenum{Sasakura_APE2013}   \\       	
  BN-flake     	&    visual       &      10.15\%       &   ----     &   \citenum{Vogl_JPD2017}   \\       	
 NV-nanocrystal 	&    AFM       &      100\%        &   ----     &   \citenum{Schroder_NL2011}   \\         	
 QD-waveguide 	&    visual      &      88\%        &   1.4\%     &   \citenum{Lee_APL2019}   \\   
  QD-$\mu$lens 	&   AM \& bulk PL       &     18\%        &   0.28\%     &   \citenum{Schlehahn_SR2018}   \\   
  QD-$\mu$lens  	&    AM \& fibre chuck      &      81.8\%        &   0.56\%     &   \citenum{Bremer_APLP2020}   \\   
  QD-cavity 	&    transmission       &      90\%        &   5\%     &   \citenum{Snijders_PRA2018}   \\   
 \hline
\hline
\end{tabular}

QD - quantum dot, NV - nitrogen vacancy, BN - boron nitride, PL - photoluminescence, AM - alignment marks, AFM - atomic force microscope, $\eta_{\mathrm{calculated}}$ - calculated efficiency assuming unity outcoupling from emitter, $\eta_{\mathrm{measured}}$ - measured single photon collection efficiency into fibre.
\end{table}
	
	
We report here on a fibre-coupled, single photon source based on a quantum dot in a bottom-up grown photonic nanowire waveguide and a compact, double gradient-index lens (GRIN) system. The bottom-up approach guarantees optimal positioning of the quantum dot within the photonic nanowire waveguide\cite{Dalacu_NT2019}, with demonstrated coupling of the dot emission to the fundamental HE$_{11}$ mode of $\beta=0.86$\cite{Reimer_PRB2016}, close to the ultimate value of $\beta \sim 0.95$\cite{Claudon_NP2010}. By appropriately tapering the photonic nanowire, a gaussian mode is transformed to free-space for perfect mode-matching to a SMF fibre\cite{Bulgarini_NL2014}. We first demonstrate efficient coupling between the nanowire source and the GRIN lens system at 4\,K in a stable, closed-cycle cryostat where the dot emission is used for alignment. We then demonstrate PnP operation by permanently fixing the GRIN lens system to the nanowire sample. In this case, we use photoluminescence from the band to band recombination in the photonic nanowire surrounding the quantum dot for aligning. Unlike the dot, emission from the host nanowire remains bright at room temperature and, importantly, the dot and nanowire are spatially co-located allowing for indirect room temperature $X$-$Y$ alignment of the source and the GRIN lens system.

	
\begin{figure}
		\includegraphics[width=0.5\textwidth]{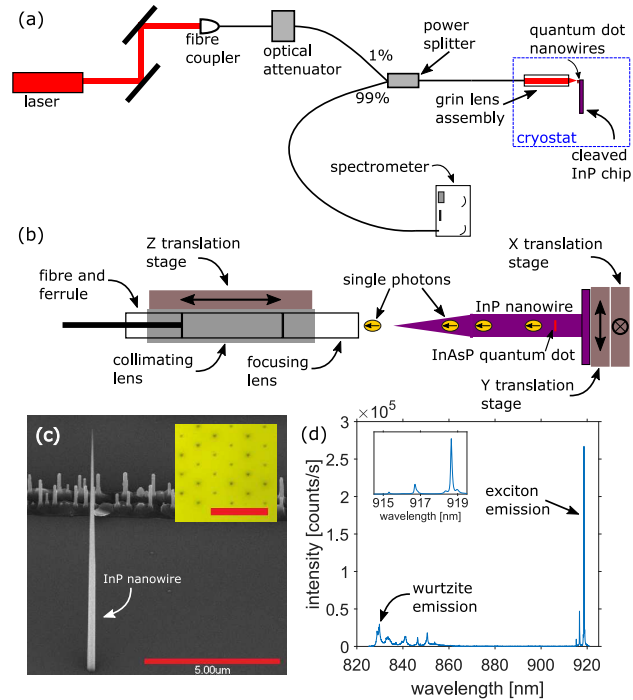}
		\caption{\label{fig:sys_overview}The optical system for detecting spectral response is shown in (a). A closer view of the GRIN lens arrangement is seen in (b). The nanowire sample can be translated in two dimensions ($X$ and $Y$), and the GRIN lens may be moved to change the focus of the lens relative to the nanowire ($Z$). The grey box overlapping the GRIN lenses and fibre ferrule is a ferrule sleeve used to maintain optical alignment. An SEM image of a nanowire can be seen in (c). Inset is a top-down optical image of the hexagonal grid of nanowires---red bar represents 20\,$\mu$m---used in subsequent scan images. The spectral plot in (d) shows a bright $X^-$ exciton emission line in a typical quantum dot used in this work. Inset in (d) is a plot capturing only the quantum dot emission.}
\end{figure}

	\begin{figure*}
		\includegraphics[width=\textwidth]{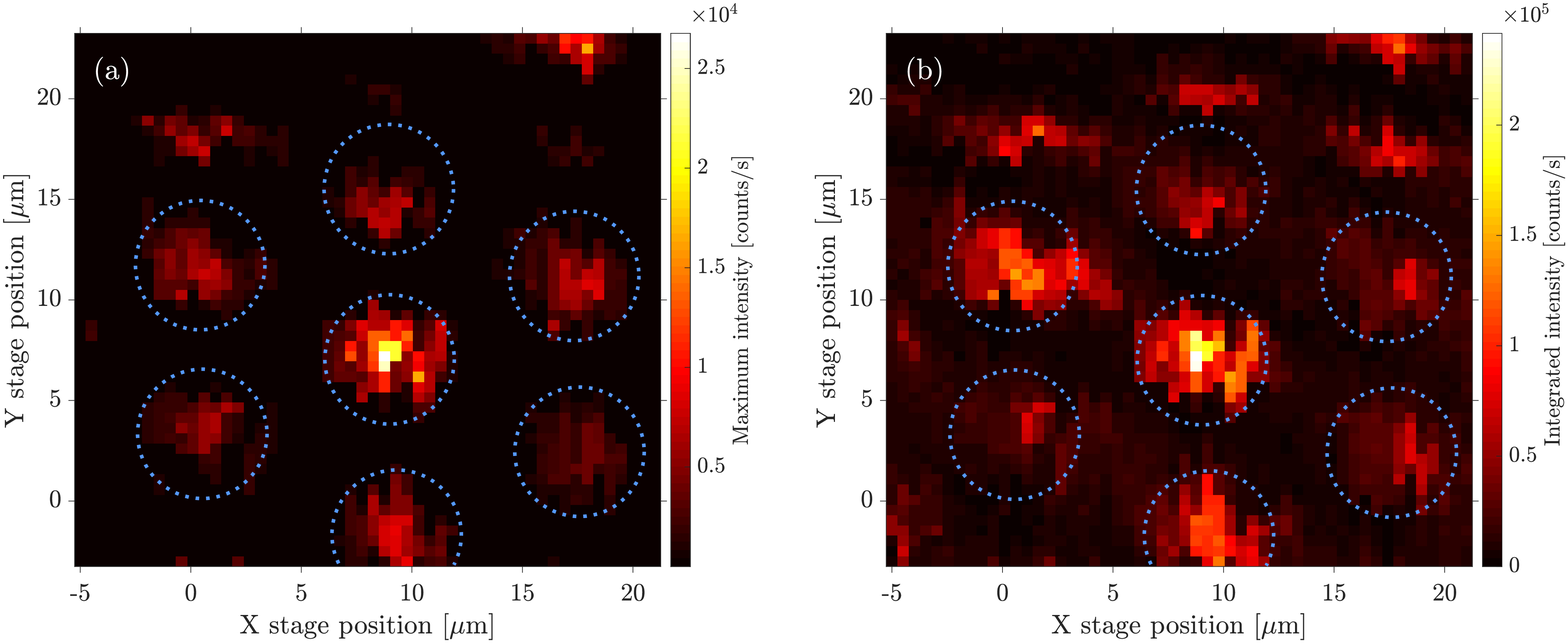}
		\caption{\label{fig:scans}Plots of quantum dot and wurtzite InP emission as a function of GRIN lens position relative to the sample substrate. (a) and (b) show data from a $25 \times 25$ $\mu$m$^{2}$ scan of a hexagonal array of nanowires (see inset image in FIG.~\ref{fig:sys_overview}(c)), with a scan step size of $0.5$ $\mu$m. The quantum dot count rate represents the maximum value in a spectral range of 908-950 nm, which is the expected range for this sample. The wurtzite InP emission in (b) is an integrated count from 823 to 836 nm. Blue dotted circles are used to emphasize the same subsection of the hexagonal array of nanowires.}
	\end{figure*}

The single photon sources used in this work are InAsP quantum dots that are embedded within InP nanowires. We use site-selective vapour-liquid-solid (VLS) expitaxy; a process described in more detail in previous studies\cite{Dalacu_NT2009,Dalacu_APL2011}. 20\,nm gold particles are lithographically defined and deposited to provide a catalyst at desired growth sites. Along with 250\,nm holes in a SiO$_2$ growth mask, the gold defines the InP nanowire diameter during the VLS epitaxy process. Nanowires are typically grown to a length of $\sim$750\,nm, and quantum dots are defined $\sim$500\,nm above the substrate. At this point, a change to the growth conditions promotes substrate growth over catalyzed growth, and the nanowire radius increases to that defined by the holes in the oxide mask. After processing, nanowires have a typical radius of $\sim$250\,nm and are $\sim$10 microns long, see Fig.~\ref{fig:sys_overview}(c). The growth method allows for the definition of a tapered tip with controlled angle, suitable for launching photons in a gaussian mode\cite{Bulgarini_NL2014} into free space, as mentioned above.

Light collection from the nanowire utilizes two GRIN lenses (Grintech GmbH) and a pig-tailed fibre (Nufern HI 1060). Emission from the nanowire is collimated with a 0.5 NA GRIN lens and re-focused into the fibre using a second lens with $\mathrm{NA}=0.2$. The schematic of the GRIN assembly is shown in FIG.~\ref{fig:sys_overview}(b). For the GRIN lens assembly, we estimate, at 950\,nm wavelength, $\sim$220\,$\mu$m for the free space focal length, and 0.76\,$\mu$m for the beam waist. This is predicted using one dimensional calculations with the transfer matrix method. Using a 965\,nm laser, a CCD camera is used to directly measure the beam waist as a function of camera-lens separation in the far field. Using $w(z) \approx \lambda(z-z_0)/(\pi n_0 w_0)$, the waist was found to be $w_0 = 0.80\,\mu$m, corresponding to a numerical aperture of $\mathrm{NA} = 0.40$. The system is designed for an NA that matches the divergence of the HE$_{11}$ mode emitted from the tapered photonic nanowire waveguide\cite{Dalacu_NM2021} in order to provide a high mode overlap and near unity collection. The lens assembly is fixed together using an optically clear epoxy (EPO-TEK 301) and a glass ferrule sleeve.
	
Low temperature photoluminescence spectra of the quantum dot sample were take with the GRIN lenses in a closed-cycle He cryostat using the system arrangement seen in FIG.~\ref{fig:sys_overview}(a). The cryostat is equipped with three orthogonal piezoelectric-driven translation stages. The GRIN lens system is mounted on the $Z$ stage whilst the sample is on the $X$-$Y$ stages. The system is optically pumped using a fibre-coupled laser source connected to the 1\% arm of a 99:1 fused fibre power splitter and focused on the sample by the GRIN lenses. 	Reflected laser light and sample emission are collected by the lens assembly and directed along the 99\% arm of the power splitter, towards a spectrometer equipped with a CCD detector array. 
	
	
	
	\begin{figure*}
		\includegraphics[width=\textwidth]{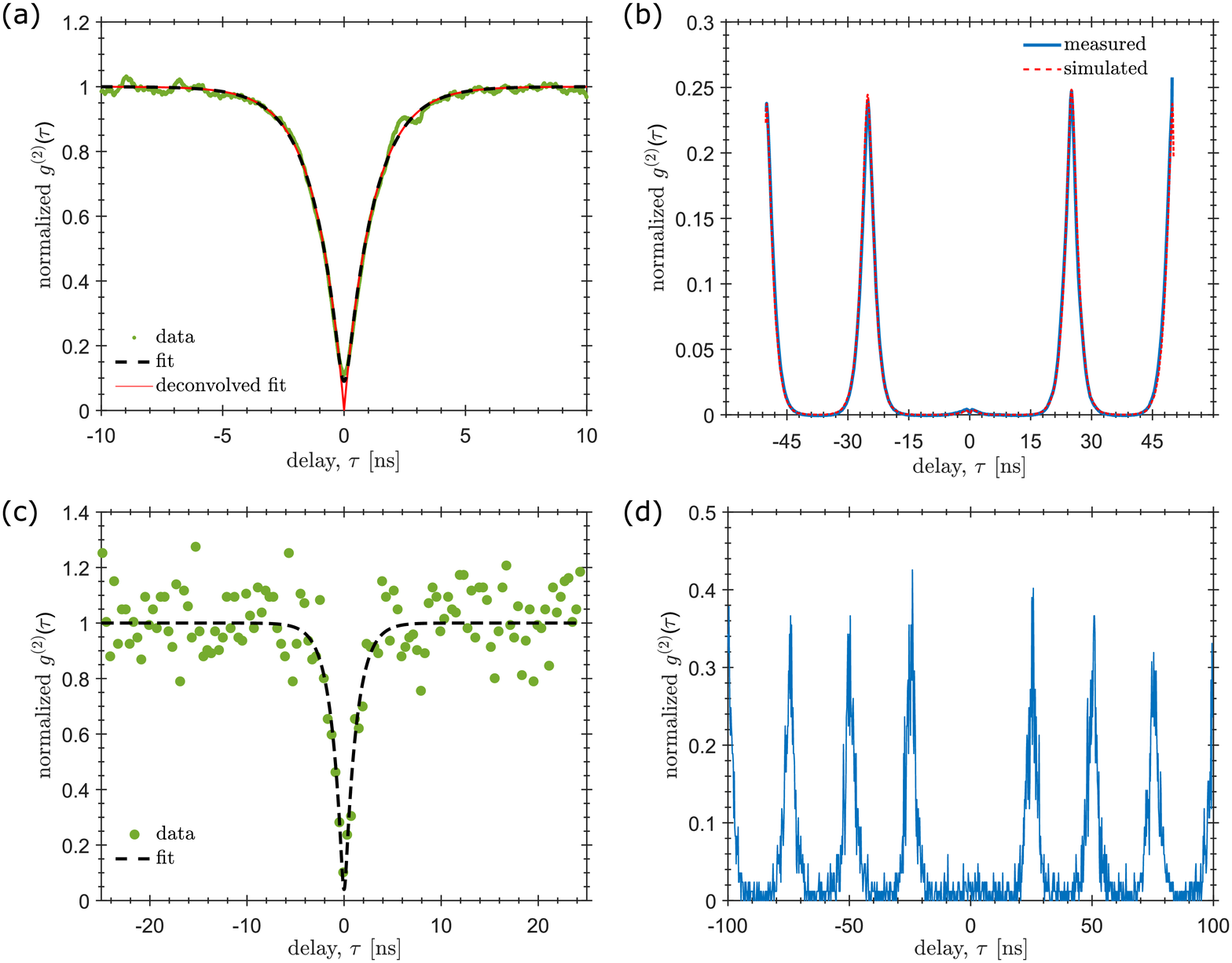}
		\caption{\label{fig:g2exp_CWplot} $g^{(2)}(\tau)$ measurements made with active [(a) and (b)] and fixed [(c) and (d)] alignment. In (a) and (c), a CW laser (HeNe, 633\,nm) is used to excite the quantum dot at saturation, with $g^{(2)}(0)=0.002$ and 0.04, respectively, after deconvolving the detector response jitter using Eq.~\ref{eq:CW_g2}. In (b) and (d), the measurements were made using pulsed excitation with a pulsed diode laser (PDL) at 670\,nm. A pulsed $g^{(2)}(\tau)=0.015$ is estimated in (b) through comparison to simulated data which is also plotted. The two sets of data (active and fixed alignment) were measured on different nanowire quantum dot devices.}
	\end{figure*}
	

	
	In lieu of conventional imaging, which is not possible using the GRIN lens assembly, we perform two dimensional (2D) scans of the sample surface using the $X$-$Y$ piezoelectric stage controllers. The scans are performed using a two stage process. We first find the focus ($Z$) of the GRIN lens assembly by maximizing the reflected power from a 930\,nm laser (similar to the dot emission wavelength). At this $Z$ position the GRIN lens assembly is focused on the substrate which is sufficient to observe emission from both the quantum dot and the InP making up the photonic nanowire. The latter adopts a wurtzite crystal phase in our nanowires, with a higher bandgap (832\,nm at 4\,K) compared to bulk InP which has a zincblende crystal structure. In the second stage, we excite above band with wavelengths of 780\,nm, 632\,nm, or 670\,nm and record the photoluminescence as a function of $X$-$Y$ position. A typical spectrum obtained when the GRIN lens assembly is over a nanowire is shown in FIG.~\ref{fig:sys_overview}(d), with both InAsP dot emission and wurtzite InP emission clearly visible.
	

 Initial scans locate the nanowires on the substrate and we use these to adjust the $Z$-focus to optimize the collection of the quantum dot emission as opposed to focusing on the substrate. Examples of such in-focus scans can be seen in FIG.~\ref{fig:scans}. The position step size used in the scans is $\sim$0.5\,$\mu$m in $X$ and $Y$, and the mode of travel of the piezoelectric drives is with slip-stick motion. Each of these coloured pixels is derived from a full spectrum measurement shown in FIG.~\ref{fig:sys_overview}(d). In FIG.~\ref{fig:scans}(a), we take the maximum count rate in the expected spectral range of the quantum dot emission (908-950\,nm) and in FIG.~\ref{fig:scans}(b) we take a sum of the counts in the spectral range 823-836\,nm which contains the peak associated with band to band recombination in the wurtzite InP nanowire. Comparison of the two figures highlights the spatial co-location of the dot and nanowire emission, important for room temperature alignment discussed later. 
	
	To determine the purity of the gathered single photons, measurements of the second order correlation function, $g^{(2)}(\tau)$ were taken in a typical Hanbury Brown and Twiss (HBT) arrangement.  We first optimize the position of the GRIN lens using the $X$-$Y$-$Z$ stages for maximum collection of emission from a chosen quantum dot. Count rates at saturation using continuous wave (CW) excitation are up to 1.4\,Mcps.  The bulk of the emission collected from the quantum dot will travel down the 99\% arm of the first power splitter. A tunable narrowband transmission filter with a $0.1$\,nm bandwidth suppresses $>45$ dB of the out-of-band transmitted signal, including the laser pump power and unwanted sample emission.  The filtered signal is fed into a 50:50 fused fibre power splitter with each 50\% arm of the splitter leading to a single-photon avalanche diode (SPAD) detector. The signals from the detectors are time-correlated using counting electronics to build a histogram of the delay times between detection events on the first and second detector, $\tau=t_2-t_1$.
	
	\begin{figure*}
		\includegraphics[width=\textwidth]{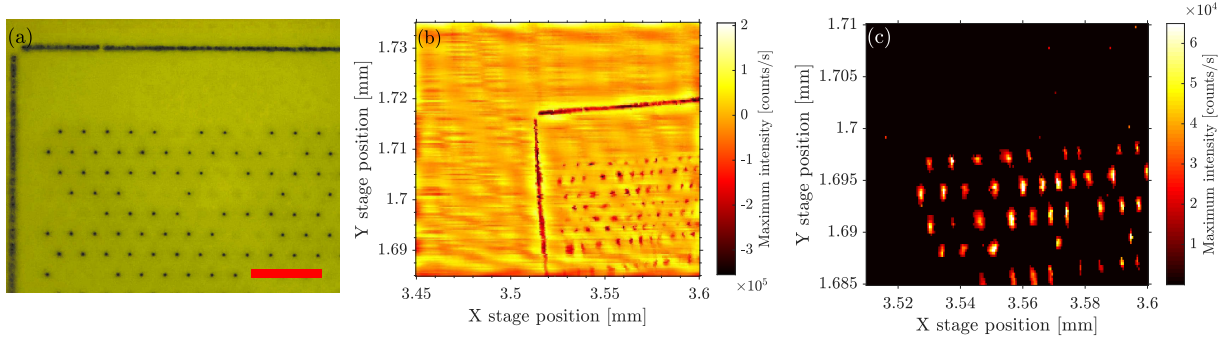}
		\caption{\label{fig:RT_images_scans} An optical image of a hexagonal array of nanowires, shown in (a), is compared with (b) a reflection (930\,nm) and (c) a wurtzite InP emission scan. The scans were performed at room temperature and atmospheric pressure. Nanowires are clearly seen and distinct in position, allowing the lens assembly to be accurately placed over any scanned nanowire. The red scale bar in (a) represents 20 $\mu$m.}
	\end{figure*}
	
	
	


	FIG~\ref{fig:g2exp_CWplot}(a) shows a CW $g^{(2)}(\tau)$ measurement made using a HeNe laser as an above-band excitation source.  Assuming the CW $g^{(2)}(\tau)$ without the effect of detectors is given by a Laplace distribution and that the detector response can be described by a gaussian, the normalized second order correlation measurement seen in the experiment is given by
	\begin{equation}
	    g^{(2)(\tau)} = 1 - \frac{A}{2}\left( B_{+} + B_{-}\right)
	    \label{eq:CW_g2}   
	\end{equation}
	where $B_{\pm}$ is defined as
	\begin{equation*}
	    B_{\pm} = \left[1 \mp erf\left( \frac{2 \tau \pm \xi^2 \Gamma}{2\xi}\right) \right] e^{\pm 4 \tau/\Gamma + \xi^2}
	\end{equation*}
	where $\Gamma$ is the radiative recombination rate of the excitonic complex, $\xi$ is the detector jitter, and $A$ describes the magnitude of the correlation minimum. The SPADs used in these measurements have a detection jitter of $\xi \approx 190$\,ps. The expected time-correlated response without the influence of detectors is also plotted against the experimental data. This curve is produced by taking the fit to the correlation data and plotting it again with $\xi = 0$.
	
	
	In FIG~\ref{fig:g2exp_CWplot}(b) we demonstrate on-demand operation using a diode laser (emitting at 670 nm) as a pulsed excitation source. Pumping at saturation using 39 ps pulses repeated at 40\,MHz, we obtain a total count rate at the detectors of 0.2\,Mcps. At this pump power, the measured $g^{(2)}_{sat}(0) = 0.015$ (i.e. the normalized area of the $\tau=0$ peak). The same value was obtained by fitting the $g^{(2)}(\tau)$ curve using a stochastic model to simulate the pulsed HBT experiment\cite{Dalacu_PRB2020}.  The high single photon purity, even at saturation and using above-band excitation, is a consequence of the deterministic growth process employed in the manufacture of the source, which produces devices containing one and only one emitter. 
	

	
	
	To estimate the device efficiency, we account for the throughput of all the fibre components (20\%) and the detector efficiency (25\%) from which we estimate 4\,Mcps single photons coupled into the fibre by the GRIN lens. Considering the excitation pulse repetition rate of 40\,MHz, and assuming the quantum efficiency for the quantum dot is 100\%, this corresponds to 10\% of the photons emitted by the dot being coupled into the fibre. To estimate the collection efficiency of the GRIN lens system itself we have to take into account the fact that not all the photons emitted by the dot will be radiated from the top of the nanowire. We assume a loss of 50\% for photons emitted by the dot that are directed down towards the substrate and 5\% into radiation modes (i.e. $\beta=95\%)$. We also have to take into account 25\% which emit from a different charge state (see Fig.~\ref{fig:sys_overview}(c)) and 20\% that emit into the phonon sidebands, both of which are filtered out. These loss channels reduce the maximum number of the desired photons at the input of the GRIN lens to 28.5\% of 40\,MHz. Accordingly, we estimate a GRIN lens assembly collection efficiency of 35\%.
	
	We speculate that the large fraction of uncollected photons (65\%) is due to imperfect alignment, in particular, a tilt of the GRIN lens system with respect to the substrate. In a separate experiment, we have measured the dependence of the coupling efficiency on the deviation of the GRIN lens axis from the substrate normal and have observed a decrease of over 50\% at angles as small as 1$^{\circ}$. Angle alignment can be readily improved by incorporating precise pitch and yaw control during the alignment process. We also note that nanowire device designs exist that can reduce some of non-alignment related sources of loss, including the incorporation of a back mirror\cite{Claudon_NP2010,Reimer_NC2012} and charge-state control\cite{Zeeshan_PRL2019}.

	We turn next to room temperature alignment and PnP operation. We focus a 930\,nm laser onto the surface looking at the reflected laser power, as before, but here, we also perform an $X$-$Y$ scan which provides information on the surface topography. An example is shown in FIG.~\ref{fig:RT_images_scans}(b) where the target array of nanowires is clearly visible. While knowing the location of nanowires using reflected laser power can give a rough stage location for GRIN lens-nanowire coupling, it is not suitable for determining if the lens is in the proper focus position to maximize quantum dot emission collection. To better position the fibre and lenses, we investigated whether room temperature wurtzite InP emission scans were possible. In FIG.~\ref{fig:RT_images_scans}(c), such a scan shows that wurtzite InP emission can clearly show distinct nanowires, allowing for precise positioning of the collection lenses at room temperature. An optical image of the same nanowire array [Fig.~\ref{fig:RT_images_scans}(a)] shows that the emission image accurately recreates the same arrangement of nanowires, where missing wires are seen in the same locations in both images.
	
	
	\begin{figure}
		\includegraphics[width=0.5\textwidth]{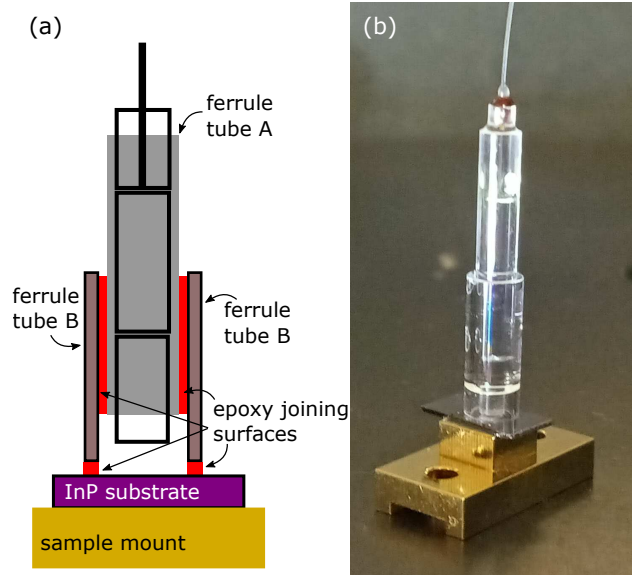}
		\caption{\label{fig:gla_fixed}The fixed alignment structure is shown in a schematic representation, (a), and in a picture of a finished device, (b). The schematic highlights the critical surfaces that are bonded together with epoxy, after alignment to a nanowire is established.}
	\end{figure}
	

	 To permanently fix the alignment of the nanowire to the fibre assembly we first fix the nanowire growth substrate onto the metal sample holder using 3M 2216 B/A epoxy adhesive. An outer ferrule tube (B in FIG.~\ref{fig:gla_fixed}(a)) is placed, loose, on top of the InP substrate, located roughly above a target array of nanowires. The GRIN lens assembly used in the low temperature measurements (A in FIG.~\ref{fig:gla_fixed}(a)) is then lowered into the outer ferrule tube. Using a 930\,nm laser, the GRIN lens assembly is brought into focus on the substrate surface using the reflected intensity to optimize – this focus position is typically a few microns below the ideal focus position for quantum dot emission. The GRIN lens assembly (A) is now glued to the inner walls of the outer ferrule tube (B) using EPO-TEK 301-2 epoxy and cured for approximately 3 hours at 75$^\circ$C. After curing, the assembly is raised a few microns and a small $X$-$Y$ scan is performed to locate a chosen nanowire using the wurtzite InP emission. Once located, the optimum focus is found and the outer ferrule tube is glued to the substrate using EPO-TEK 301-2 epoxy and further cured at 75$^\circ$C for another 3 hours. A picture of the finished device is shown in FIG.~\ref{fig:gla_fixed}(b).
	
	
	The device was then cooled to 4\,K and operated as a PnP source. Measured count rates were $\sim 10^4$\,cps, significantly lower compared to devices employing active alignment at low temperature, but single photon purity remained high, with $g^{(2)}(0) \sim 0.01$, see FIG.~\ref{fig:g2exp_CWplot}(c), (d). This initial device was intended to verify the feasibility of the approach, and, in particular, the alignment stability. No variation in count rate was observed after several cool-downs in the cryostat, nor was any observed after several immersions in liquid N$_2$. 
	
	To increase the count rate to the levels achieved using active low temperature alignment, more careful attention to the chosen focus position at which the GRIN lens is fixed is required. In a separate experiment we found that the focus position for optimal collection of the InP emission occurs a few microns below that of the quantum dot emission. This misalignment in the focus that occurs with room temperature InP emission alignment can be readily calibrated out.	 Another interesting avenue is to apply the technique to telecom nanowire quantum dot emitters\cite{Haffouz_NL2018}. The quantum dots used in these sources have a deeper confining potential and emit brightly even at room temperature\cite{Fiset-Cyr_APL2018}. Consequently, emission from the dots themselves can be used for alignment instead of the InP nanowire emission.
     
    In summary, we demonstrate a process to efficiently couple single photons from InAsP quantum dots embedded in InP photonic nanowire waveguides into a single mode fibre using a compact and robust GRIN lens assembly. The collection efficiency of the GRIN lens system (e.g. fraction of the single photons directed towards the lens assembly that were collected) was 35\% whilst the overall single photon collection efficiency (fraction of the total single photons produced by the quantum dot that were collected) was 10\%. We also describe a manufacturing method to produce high purity single photon sources that can be operated in a ``plug and play" mode. The method is based on active alignment using the emission of the InP nanowire in which the InAsP quantum dot is embedded. This alignment can be performed at room temperature, where the fibre can be permanently attached to the chip. 
	
	\begin{acknowledgments}
		This work was supported by the Canadian Space Agency through a collaborative project entitled `Development of Quantum Dot Based QKD-relevant Light Sources'
	\end{acknowledgments}

	\bibliography{whiskers}
	
\end{document}